# Canonical Reduction of Tensors and the Physical Properties of Condensed Matter: Application to Optics

R. BONNEVILLE

*Centre National d'Études Spatiales, 2 place Maurice Quentin, 75001 Paris, France*

A general framework for the description of the physical properties of matter by a canonical reduction procedure of tensors is presented; besides geometrical symmetries, this paper emphasizes the role of intrinsic symmetries which are due either to the indiscernability of some of the physical quantities involved or to thermodynamical arguments. The intrinsic symmetries are expressed through the behaviour of the tensors describing the investigated property under the effect of some index permutation. The scheme of reduction of any tensor into parts that are irreducible not only with respect to rotations and inversion but also with respect to index permutations is shown and examples are given in the area of light-matter interaction.

## INTRODUCTION

The interest in using an irreducible tensor formalism for describing the physical properties of condensed matter is well known. With such a formalism, for instance, there is easy writing of the selection rules imposed by the geometry of the system as well as the simplicity of performing such geometrical operations as the transformation from a microscopic frame to a macroscopic one. The reduction of a cartesian tensor into parts that are irreducible with respect to the orthogonal group or to a particular point-group, either in the cartesian or in the spherical formalism, has been systematically reviewed and many examples of the applications of that approach have been given (for examples see Reference 1 and bibliography therein).

Nevertheless, in most cases the decomposition of a rank-r cartesian tensor $T_{ijk\ldots}$ is not unique; alternative reduction schemes are possible whenever one can find in the decomposition results several linearly independent parts belonging to the same irreducible representation of the orthogonal group. For instance a rank-3 tensor $T_{ijk}$ has three mutually orthogonal vector parts; given such a triplet, any other tensor deduced from the former through a unitary transformation is also convenient. One thus may ask what is the best reduction scheme, i.e. the most suitable for a given physical problem; the aim of this paper is to clarify this point.

Any tensor may undergo two types of transformations; firstly, it may change the value (i.e. *x* or *y* or *z)* of the cartesian indices independently of their position (geometrical transformation); secondly, it may change the position of the cartesian indices independently of their value (intrinsic symmetry). Most often, physical effects have intrinsic symmetries which proceed either from the indiscernability of some among the quantities involved or from thermodynamical arguments, and the manifestation of those intrinsic symmetries is to affect the behaviour of the tensors associated with certain physical quantities under a class of index permutations. For instance, the second order electromagnetic susceptibility tensor $\chi^{(2)}_{ijk}(-2\omega,\omega,\omega)$ describing second-harmonic generation is symmetric in the *j*-*k* permutation because of the indiscernability of the two fundamental fields; the third rank tensor $\eta_{ijk}$ which accounts for optical activity is antisymmetric in the permutation of its extreme indices *i* and *k* from a general property of the kinetic coefficients.

This gives the key of the problem: before decomposing a tensor into parts irreducible under geometrical operations, one first has to decompose it into parts irreducible under index permutations and to extract those which are compatible with the intrinsic symmetries of the studied effect[2]; each of these latter parts can then be reduced under the three-dimension orthogonal group and ultimately under the particular point group of the system under investigation[3].

Let $T$ be any rank-r tensor; $T$ will split under the permutation group of r objects $\sigma(\mathrm{r})$ into a set of linearly independent rank-r tensors $t^{\{\Gamma\}_1}, t^{\{\Gamma\}_2}, ..., t^{\{\Gamma\}_q}$ ; $t^{\{\Gamma\}_q}$ has the same transformation properties under $\sigma(\mathrm{r})$ as the q$^{th}$ base vector $\{\Gamma\}_q$ of the representation $\{\Gamma\}$ of $\sigma(\mathrm{r})$; then each of the $t^{\{\Gamma\}_q}$ s will in turn split into a set of irreducible tensors $t^{\{\Gamma\}_{q,\mathrm{J}\pi}}$ possessing the symmetry $\{\Gamma\}_q$ and belonging to the irreducible (2J + 1)-dimensional representation $D^{\mathrm{J}\pi}$ of weight J and parity π of the three-dimension orthogonal group O(3). Finally the following equation constitutes the reduction spectrum of $T$:

$$T = \sum_{\Gamma,q} \sum_{J,\pi} t^{\{\Gamma\}_{q,\mathrm{J}\pi}}$$

In the two following parts of this paper, we will present in more detail the two steps of this approach. The two last parts treat specific examples in the area of light-matter interaction: linear susceptibility, optical activity and magneto-optical effects, and nonlinear optical processes.

## REDUCTION PROCESS, PERMUTATIONS

Whereas handling the irreducible representations of the three-dimension orthogonal group as well as those of the point groups is in current use in condensed matter physics, the properties of the irreducible representations of the permutation groups are far less well known and most often are only referred to in studies of the exchange of identical particles. The main results, which will be useful for the present study, are listed hereafter; for more details, the reader is referred, for instance, to References 4 and 5.

Considering the (non-commutative) permutation group of r objects $\sigma(\mathrm{r})$, any of the r! elements of $\sigma(\mathrm{r})$ can be expressed as a product of pair permutations; all the properties of $\sigma(\mathrm{r})$ can thus be derived from those of $\sigma(2)$. It can be shown that there is a one-to-one correspondence between the irreducible representations $\{\Gamma\}$ of $\sigma(\mathrm{r})$ and the so-called partitions of r*:* a partition of r is a set of integers $\left\{n_,, n_2, n_3, ..., n_q\right\}$ with $n_1 \geq n_2 \geq n_3 \geq ... \geq n_q$ and $n_1 + n_2 + n_3 + ... + n_q = r$. For instance, given the permutation group of four objects $\sigma(4)$, five partitions of 4 are possible,

$\{4\}, \{3,1\}, \{2,2\}, \{2,1,1\}, \{1,1,1,1\}$; $\sigma(4)$ hence has five irreducible representations which are respectively noted as

$\{4\}, \{31\}, \{2^2\}, \{21^2\}, \{1^4\}$ where the number of occurrences of a given integer has been put as an exponent. The irreducible representation associated with the partition $\left\{n_,, n_2, n_3, ..., n_q\right\}$ is graphically shown by a-so-called Young array, with $n_1$ squares on the first line, $n_2$ squares on the second line (starting from the left), and so on. The base vectors of the representation are figured by filling the r squares of the array with the r objects in a proper arrangement; this arrangement can be derived through the following recursive process. One starts with two objects: $\sigma(2)$ has two irreducible representations, a symmetric one $\{2\}$ and an antisymmetric one $\{1^2\}$, both of dimension 1 (Figure 1(a)). The base vectors of the irreducible representations of $\sigma(3)$ are merely obtained by adding a third object wherever doing so generates a possible Young array for $\sigma(3)$, as shown in Figure 1(b). Thus are generated, (i) one single vector belonging to the representation $\{3\}$, symmetric in the exchange of any pair of indices; (ii) two linearly independent vectors belonging to the representation $\{21\}$, symmetric in the exchange of a pair of indices and antisymmetric in the exchange of another pair; (iii) one single vector belonging to the representation $\{1^3\}$, antisymmetric in the exchange of any pair of indices. This process is straightforwardly extended to $\sigma(4)$ and beyond (Figure 1(c)).

It appears as a consequence that the dimension of any irreducible representation of $\sigma(\mathrm{r}+1)$ corresponding to a given partition of (r + 1) is the sum of the dimensions of all the irreducible representations of $\sigma(\mathrm{r})$ corresponding to partitions of r generating that partition by the addition of a (r + 1)$^{\mathrm{th}}$ object. In Figure 2, the dimensions $\mathrm{d}(\Gamma)$ of the irreducible representations $\{\Gamma\}$ of $\sigma(\mathrm{r})$ have been reported for r = 2, 3, 4; one can check that $\sum_{\Gamma} \mathrm{d}(\Gamma)$ is equal to n!, i.e. the number of elements of $\sigma(\mathrm{r})$. One notes that the base vector figured by a given Young array is symmetric in the exchange of two indices on the same line and antisymmetric in the exchange of two indices on the same column; moreover $\sigma(\mathrm{r})$ always has one single fully symmetric representation $\{\mathrm{r}\}$ and one single fully antisymmetric representation $\{1^{\mathrm{r}}\}$, both of dimension 1.

One can now return to the original problem of splitting a tensor into irreducible parts.

Given any group G of transformations within a N-dimension vector space $E$ , the rank-r tensors subtend a tensor space $E \otimes E \otimes E \otimes ... \otimes E = \otimes^N E$ , which may be viewed as a $N^r$-dimension reducible representation of G. A general way of reducing any tensor $T$ of this type would consist of first reducing $T$ with respect to a supergroup of G. Here, G will be the orthogonal group O(N) and as a supergroup we will take the full linear group GL(N). It can be shown that the only tensor operations which commute with GL(N) are the index permutations. The tensors which belong to a given irreducible representation of the permutation group $\sigma(r)$ belong to the same irreducible representation of GL(N). The tensor space $\otimes^N E$ is therefore reducible with respect to GL(N) into tensor subspaces which are irreducible under $\sigma(r)$; however, those subspaces are not irreducible under the orthogonal group O(N).

The reducible $N^r$ -dimension space subtended by the rank-r tensors splits into irreducible subspaces whose dimensions are easy to find, as will be shown below. Firstly, the dimension of the tensor subspace possessing the permutation symmetry of the fully symmetric representation $\{r\}$ is the number of ways of giving a value to the r indices, with an allowed repetition, i.e. $\binom{N+r-1}{r}$. Secondly the dimension of the tensor subspace having the permutation symmetry of the fully antisymmetric representation $\{1^r\}$ is the number of ways of giving a value to the *r* indices with no repetition allowed, i.e. $\binom{N}{r}$ if r <N, 0 otherwise. In Figure 3 the recursive generating process of the base vectors of the irreducible representations of $\sigma(r)$ has been reprinted from Figure 1 but in addition under each base vector of $\sigma(r)$ the number of independent components of a tensor transforming in the same way as the base vector under $\sigma(r)$ has been indicated, either explicitly for the tensor subspaces of respective symmetries $\{\mathrm{r}\}$ and $\{1^{\mathrm{r}}\}$, or formally. From the relations between these numbers, it is easy to calculate, step by step, the explicit number of independent components for all the tensors irreducible under index permutations.

## REDUCTION PROCESS, ORTHOGONAL GROUP

Henceforth N= 3 and each index *i, j, k, ...* will be given any of the three values *x, y, z.* The number of independent components of a rank-2, 3, 4 tensor irreducible under index permutations is shown in Figure 4. With the help of Figure 4 and using the composition rule of the irreducible representations of O(3), i.e.

$$D^{\mathrm{J}\pi} \otimes D^{\mathrm{J}'\pi'} = \sum_{\mathrm{K}=|\mathrm{J}-\mathrm{J}'|}^{\mathrm{J}+\mathrm{J}'} D^{\mathrm{K}\pi.\pi'}$$

one can obtain the reduction spectrum of any cartesian tensor. This recursive reduction process is summarized in Figure 5, where a spectroscopic notation has been adopted for the irreducible tensor parts: an irreducible tensor part of weight J = 0, 1, 2, 3, 4, …, transforming like the base vector $\{\Gamma\}_q$ under index permutations, will be labelled by $\{\Gamma\}_q$ followed by a capital letter such as $\mathrm{S}, \mathrm{P}, \mathrm{D}, \mathrm{F}, \mathrm{G}, \ldots$ . for a true tensor (parity $\pi=(-1)^{\mathrm{J}}$) and by an overlined capital letter $\overline{\mathrm{S}}, \overline{\mathrm{P}}, \overline{\mathrm{D}}, \overline{\mathrm{F}}, \overline{\mathrm{G}}, \ldots$ for a pseudo-tensor (parity $\pi=(-1)^{\mathrm{J}+1}$). It has been assumed that the original rank-r cartesian tensor $T$ is a true tensor (parity $(-1)^{\mathrm{r}}$); if $T$ is a pseudo-tensor (parity $(-1)^{\mathrm{r}+1}$), the representations S and $\overline{\mathrm{S}}$, P and $\overline{\mathrm{P}}$, D and $\overline{\mathrm{D}}$, F and $\overline{\mathrm{F}}$, ... respectively must be exchanged.

In order to obtain the explicit expressions of the irreducible components of a rank-r tensor $T$ in terms of the original cartesian ones $T_{ijk\,\ldots}$ one must first reduce $T$ under $\sigma(\mathrm{r})$ by projecting it upon the base vectors $\{\Gamma\}_q$ of the relevant irreducible representations $\{\Gamma\}_q$ of $\sigma(\mathrm{r})$ and so build a set of linearly independent cartesian tensors $t^{\{\Gamma\}_q}$ transforming like $\{\Gamma\}_q$ under index permutations. This will be performed by applying the projection operators whose explicit expressions can be obtained by remembering that the eigenvector represented by a given Young array is symmetric in the permutation of any pair of indices on the same line and antisymmetric in the exchange of any pair of indices in the same column. I being the identity operator and $\hat{P}(1,\ 2)$ the permutation operator of two objects, 1 and 2, the symmetrizing operator of any function of the pair (1, 2) is

$$\hat{S}(1,2) = \left(I + \hat{P}(1,2)\right)/2\,,$$

and the anti-symmetrizing operator is

$$\hat{A}(1,2) = \left(I - \hat{P}(1,2)\right)/2\,.$$

More generally, the symmetrizer of any function of q objects (1,2, ...,q) is given by

$$\hat{S}(1,2,...,\mathrm{q}) = \sum_P \hat{P}(1,2,...,\mathrm{q})/\mathrm{q}!$$

where the sum runs over the q! different permutations $\hat{P}(1,2,...,\mathrm{q})$ of (1,2, ...,q), and the anti-symmetrizer by

$$\hat{A}(1,2,...,\mathrm{q}) = \sum_P \varepsilon(\hat{P})\hat{P}(1,2,...,\mathrm{q})/\mathrm{q}!$$

with $\varepsilon(\hat{P}) = +1$ (respectively (–1)) if $\hat{P}$ writes as the product of an even (odd) number of pair permutations.

Let one now consider the Young array associated with a given base vector $\{\Gamma\}_q$ of a d-dimension irreducible representation of $\sigma(\mathrm{r})$ and the symmetrizers $\hat{S}(1,2,...,\mathrm{q}_l)$ of the lines *(l)* of the array and the anti-symmetrizers $\hat{A}(1,2,...,q_c)$ of the columns *(c);* $\hat{S}(1,2,...,q_l)$ and $\hat{A}(1,2,...,q_c)$ act on the indices which fill the squares of the array. The operator

$$\hat{P}/\{\Gamma\}_q = (d/r!)\prod_c q_c!A(1,2,...,q_c)\prod_l q_l!\hat{S}(1,2,...,q_l)$$

where the products run over all the lines and all the columns of the array, is the projector upon the selected base vector. The d projectors associated with the d Young arrays featuring the d base vectors of $\{\Gamma\}$ thus provide a complete set of base vectors for that representation. Note that another, different but equivalent, set of base vectors could be obtained by the application of the projecting operators

$$(d/r!)\prod_l q_l!\hat{S}(1,2,...,q_l)\prod_c q_c!\hat{A}(1,2,...,q_c)$$

Remembering that $\sum_\Gamma d(\Gamma) = n!$, one can check that the sum of the projectors upon all the base vectors of all the irreducible representations of $\sigma(r)$ is just the identity. Every tensor $t^{\{\Gamma\}_q}$ generated as above will then be reduced under O(3) by projecting it upon the irreducible representations $D^{J\pi}$ of O(3); after noticing that a given representation $D^{J\pi}$ will appear at most only once in the projection of a given $t^{\{\Gamma\}_q}$, that reduction can actually be done by purely cartesian techniques by contracting $t^{\{\Gamma\}_q}$ either with the symmetric scalar unit tensor $\delta_{ij}$ or with the antisymmetric pseudo-scalar unit tensor $\varepsilon_{ijk}$. Examples for rank-2 and rank-3 cartesian tensors will be given below.

The reduction spectrum of a cartesian rank-2 true tensor is $\{2\}(S + D)$, $\{1^2\}P$. The projector upon $\{2\}$ is the symmetrizer $\left(I + \hat{P}(1,2)\right)/2$, hence

$$t^{\{2\}}{}_{ij} = \left(T_{ij} + T_{ji}\right)/2$$

$$\{2\}\mathrm{S}_{ij} = \left(\sum_{pq} \delta_{pq} t^{\{2\}}{}_{pq} / 3\right)\delta_{ij}$$
$$= \left(T_{xx} + T_{yy} + T_{zz}\right)\delta_{ij} / 3$$

$$\{2\}\mathrm{D}_{ij} = t^{\{2\}}{}_{ij} - \{2\}\mathrm{S}_{ij}$$
$$= \sum_{pq} \left(\delta_{ip}\delta_{jq} - \delta_{ij}\delta_{pq} / 3\right) t^{\{2\}}{}_{pq}$$
$$= \sum_{pq} \left(3\delta_{ip}\delta_{jq} - \delta_{ij}\delta_{pq}\right)\left(T_{pq} + T_{qp}\right)/6$$

The projector upon $\left\{1^2\right\}$ is the anti-symmetrizer

$\left(I - \hat{P}(1,2)\right)/2$, hence

$$t^{\{1^2\}}{}_{ij} = \left(T_{ij} - T_{ji}\right)/2 = \{1^2\}\overline{\mathrm{P}}_{ij} = \sum_{k} \varepsilon_{ijk} \sum_{pq} \varepsilon_{kpq} T_{pq}$$

The reduction spectrum of a rank-3 cartesian true tensor is

$$\{3\}(\mathrm{P}+\mathrm{F}), \{21\}2\left(\mathrm{P}+\overline{\mathrm{D}}\right), \left\{1^3\right\}\overline{\mathrm{S}}$$

The projector upon {3} is

$$(1/6)\hat{S}(i,\ j,\ k)$$
$$= \left(I + \hat{P}(i,\ j) + \hat{P}(j,\ k) + \hat{P}(k,\ i) + \hat{P}(i,k)\hat{P}(k,j)\hat{P}(i,j)\ \hat{P}(j,\ k)\right)/6$$

hence

$$t^{\{3\}}{}_{ijk} = \left(T_{ijk} + T_{jik} + T_{ikj} + T_{kji} + T_{kij} + T_{jki}\right)/6$$

One derives

$$\{3\}\mathrm{P}_{ijk} = \sum_{qr}\left(\delta_{qr} t^{\{3\}}{}_{iqr}\right)\delta_{jk}$$

$$= \sum_{pqr}\left(\delta_{ip}\delta_{qr}\delta_{jk}/3\right)t^{\{3\}}{}_{pqr}$$

$$\{3\}\mathrm{F}_{ijk} = t^{\{3\}}{}_{ijk} - \{3\}\mathrm{P}_{ijk}$$

$$= \sum_{pqr}\delta_{ip}\left(3\delta_{jq}\delta_{kr} - \delta_{jk}\delta_{qr}\right)t^{\{3\}}{}_{pqr}$$

The projector upon $\left\{1^3\right\}$ is

$$(1/6)\,\hat{A}(i,\, j,\, k)$$
$$= \left(I - \hat{P}(i,\, j) - \hat{P}(j,\, k) - \hat{P}(k,\, i) + \hat{P}(i,k)\,\hat{P}(k,j) + \hat{P}(i,j)\ \hat{P}(j,\, k)\right)/6$$

Hence

$$t^{\{1^3\}}{}_{ijk} = \left(T_{ijk} - T_{jik} - T_{ikj} - T_{kji} + T_{kij} + T_{jki}\right)/6$$

$$= \{1^3\}\mathrm{S}_{ijk} = \left(\sum_{pqr}\varepsilon_{pqr}T_{pqr}\right)\varepsilon_{ijk}$$

The representation $\{21\}$ has two base vectors which we will write $\{21\}'$ and $\{21\}''$; a suitable set of base vectors for $\{21\}$ is obtained through the action of the projectors

$$(1/3)\,\hat{A}(i,\, k)\ \hat{S}(i,\, j) = (1/3)\left(I - \hat{P}(i,\, k) + \hat{P}(i,\, j) - \hat{P}(i,\, k)\ \hat{P}(i,\, j)\right)$$

for $\{21\}'$ and

$$(1/3)\,\hat{A}(i,j)\,\hat{S}(i,\, k) = (1/3)\left(I - \hat{P}(i,\, j) + \hat{P}(i,\, k) - \hat{P}(i,\, j)\ \hat{P}(i,\, k)\right)$$

for $\{21\}''$

Hence the corresponding base vectors are

$$t^{\{21\}'}{}_{ijk} = \left(T_{ijk} - T_{kji} + T_{jik} - T_{jki}\right)/3$$

$$t^{\{21\}''}{}_{ijk} = \left(T_{ijk} - T_{jik} + T_{kji} - T_{kij}\right)/3$$

One derives

$$\{21\}'\mathrm{P}_{ijk} = \sum_{pqr}\left(\delta_{ip}\delta_{jk}\delta_{qr}/3\right)t^{\{21\}'}{}_{pqr}$$

$$\{21\}''\mathrm{P}_{ijk} = \sum_{pqr}\left(\delta_{ip}\delta_{jk}\delta_{qr}/3\right)t^{\{21\}''}{}_{pqr}$$

$$\{21\}'\overline{\mathrm{D}}_{ijk} = \sum_{pqr}\left(\varepsilon_{ipq}\varepsilon_{rjk}/2\right)t^{\{21\}'}{}_{pqr}$$
$$= \sum_{pqr}\delta_{ip}\left(3\delta_{jq}\delta_{kr} - \delta_{jk}\delta_{qr}\right)t^{\{21\}'}{}_{pqr}$$

$$\{21\}''\overline{\mathrm{D}}_{ijk} = \sum_{pqr}\left(\varepsilon_{ipq}\varepsilon_{rjk}/2\right)t^{\{21\}''}{}_{pqr}$$
$$= \sum_{pqr}\delta_{ip}\left(3\delta_{jq}\delta_{kr} - \delta_{jk}\delta_{qr}\right)t^{\{21\}''}{}_{pqr}$$

Now, a physical quantity such as a scattering amplitude U writes as a scalar product between two tensors of the same rank, one of them, $\chi$, being a susceptibility of the medium and the other one, $\xi$, containing the applied excitations:

$$\mathrm{U} = \sum_{ijk\ldots}\chi_{ijk\ldots}\ \xi_{ijk\ldots}$$

This scalar product can be expressed as a sum of scalar products of irreducible tensors:

$$\mathrm{U} = \sum_{\Gamma,q}\sum_{J\pi}\chi^{\{\Gamma\}q,\mathrm{J}\pi}\cdot\xi^{\{\Gamma\}q,\mathrm{J}\pi}$$

The explicit calculation of this scattering amplitude can be achieved through the cartesian formalism in a straightforward but rather tedious way and it appears somehow simpler for high rank tensors to use a spherical formalism. The methods given in Reference 1, which require the Clebsh—Gordan coefficients algebra, are cumbersome and a simpler technique of reduction of any rank-r cartesian tensor $T$ will be presented now. The reduction spectrum of $T$ has been previously determined (Figure 5); a given irreducible part of $T$ labelled as $t^{\{\Gamma\}q,J\pi}$ has a weight J and a parity π and transforms like the q^th base vector of the irreducible representation $\{\Gamma\}$ of the permutation group $\sigma(\mathrm{r})$.

One can first consider the case where $t^{\{\Gamma\}q,J\pi}$ is a true tensor (parity $(-1)^{r}$). The reduced spherical harmonics $C_M^J = \sqrt{\dfrac{4\pi}{2J+1}}\; Y_M^J(\theta,\varphi)$ form an orthogonal base of the irreducible representation of the rotation group SO(3). Let $u_x = \sin\theta\cos\varphi,\ u_y = \sin\theta\sin\varphi,\ u_z = \cos\theta$ be the coordinates of a unit vector in the direction of polar angles $\theta$ and $\varphi$. From the explicit expression of the $C_M^J$ (see Table 1), and using $u_x u_x + u_y u_y + u_z u_z = 1$, it is a straightforward matter to build an homogeneous rank-r polynomial in $u_x,\ u_y, u_z$ transforming like $C_M^J$ under any rotation; one notes that $C_M^J * = (-1)^M C_{-M}^J$, and that $\sum_M C_M^J C_M^J * = 1$. By applying to that polynomial the projector $\hat{P}/\{\Gamma\}_q$ one generates a homogeneous polynomial having the required transformation properties under both permutations and rotations. The expression of the latter polynomial provides the expression of the irreducible tensor component having the same properties.

Next, one can consider the case where $t^{\{\Gamma\}q,J\pi}$ is a pseudo-tensor (parity $(-1)^{r+1}$). From the expression of the $C_M^J$ one can build an homogeneous rank-(r–1) polynomial in $u_x,\ u_y,\ u_z$ transforming like $C_M^J$ under any rotation; then one can make, for a single given index position (arbitrarily chosen but identical for all the terms of the polynomial), the following substitution: $u_x \to u_y u_z - u_z u_y, u_y \to u_z u_x - u_x u_z, u_z \to u_x u_y - u_y u_x$ ; the resulting rank-r homogeneous polynomial behaves like a pseudo-tensor of weight J and henceforth will be noted as $\overline{C_M^J}$ . Finally, one applies to the latter polynomial the projector $\hat{P}/\{\Gamma\}_q$ as in the true tensor case.

An example of that process will now be given for the $0^{th}$ irreducible components of a cartesian rank-2 and a rank-3 true tensor. The resulting expressions are given at a normalization factor which will be commented upon later. The reduction spectrum of a cartesian rank-2 true tensor is $\{2\}(\mathrm{S}+\mathrm{D}),\ \{1^2\}\overline{\mathrm{P}}$ . By comparison with the naturally symmetrized expressions $C_0^0 = u_x u_x + u_y u_y + u_z u_z$ and $C_0^2 = 2u_z u_z - u_x u_x - u_y u_y$ it can be found straightforwardly that

$$\{2\}\mathrm{S} \propto T_{xx} + T_{yy} + T_{zz}$$

$$\{2\}\mathrm{D}_0 \propto 2T_{zz} - T_{xx} - T_{yy}$$

and from $C_0^1 = u_z$, which is readily changed into the naturally antisymmetrized expression $\overline{C_0^1} = u_x u_y - u_y u_x$ , it can be found that

$$\{1^2\}\overline{\mathrm{P}}_0 = T_{xy} - T_{yx}.$$

The reduction spectrum of a rank-3 cartesian true tensor is

$$\{3\}(\mathrm{P}+\mathrm{F}),\ \{21\}2\left(\mathrm{P}+\overline{\mathrm{D}}\right),\ \{1^3\}\overline{\mathrm{S}}.$$

From $C_0^1 = u_z = u_z\left(u_x u_x + u_y u_y + u_z u_z\right)$ and $C_0^3 = \left(2u_z u_z u_z - 3u_z u_x u_x - 3u_z u_y u_y\right)/2$ and by application of the projection operator $\hat{P}/\{3\}$ , one can find

$$\{3\}\,\mathrm{P}(0) \propto T_{xzx} + T_{zxx} + T_{xxz} + T_{yzy} + T_{zyy} + T_{yyz} + 3T_{zzz}$$

$$\{3\}\,\mathrm{F}(0) \propto -T_{xzx} - T_{zxx} - T_{xxz} - T_{yzy} - T_{zyy} - T_{yyz} + 2T_{zzz}$$

As for the representation $\{1^3\}$, the comparison with $C_0^0 = 1 = u_x u_x + u_y u_y + u_z u_z$ which can be changed into the naturally fully anti-symmetrized expression

$\overline{C_0^0} = (u_x u_y u_z - u_x u_z u_y + u_y u_z u_x - u_y u_x u_z + u_z u_x u_y - u_z u_y u_x)$ yields

$$\{1^3\}\overline{\mathrm{S}} \propto T_{xyz} - T_{xzy} + T_{yzx} - T_{yxz} + T_{zxy} - T_{zyx}$$

From $C_0^1 = u_z = u_z\left(u_x u_x + u_y u_y + u_z u_z\right)$ and by application of the projection operators $\hat{P}/\{21\}'$ and $\hat{P}/\{21\}''$ one can derive

$$\{21\}'\mathrm{P}(0) \propto T_{zxx} - 2T_{xxz} + T_{xzx} + T_{zyy} - 2T_{yyz} + T_{yzy}$$
$$\{21\}''\mathrm{P}\ \{0\} \propto T_{zxx} - 2T_{xzx} + T_{xxz} + T_{zyy} - 2T_{yzy} + T_{yyz}$$

From $C_0^2 = 2u_z u_z - u_x u_x - u_y u_y$, which can be changed into

$\overline{C_0^2} = (2u_z u_x u_y - 2u_z u_y u_x - u_x u_y u_z + u_x u_z u_y - u_y u_z u_x + u_y u_x u_x)/2$ and by application of the projectors one can find

$$\{21\}'\overline{\mathrm{D}}(0) \propto T_{zxy} - T_{zyx} + T_{xzy} - T_{yzx}$$

$$\{21\}''\overline{\mathrm{D}}(0) \propto T_{zxy} - T_{zyx} + T_{yxz} - T_{xyz}$$

The above expressions of the irreducible spherical tensor components in terms of linear combinations of the cartesian components are not normalized, while the transformation from cartesian to spherical coordinates is a unitary one. Let $\tau$ be any set of indices which allow one to discriminate among the various irreducible parts with the same weight J*;* taking normalization into account leads to $t_M^{\tau J} = \sum_{ijk...} \langle \tau JM \,|ijk...\rangle T_{ijk...}$ , with $\sum_{ijk...} \left|\langle \tau JM \,|ijk...\rangle\right|^2 = 1$. The results are as follows (N.B.

the phases of the coupling coefficients $\langle \tau JM \,|ijk...\rangle$ have been chosen to be $i^{(r-J)}$ so as to be consistent with Reference I):

$$\{2\}\mathrm{S} = \left(-1/\sqrt{3}\right)\left(T_{xx} + T_{yy} + T_{zz}\right)$$

$$\{2\}\mathrm{D}(0) = \left(+1/\sqrt{6}\right)\left(2T_{zz} - T_{xx} - T_{yy}\right)$$

$$\{1^2\}\overline{\mathrm{P}}(0) = \left(i/\sqrt{2}\right)\left(T_{xy} - T_{yx}\right)$$

$$\{3\}\mathrm{F}(0) = \left(1/\sqrt{10}\right)\left(-T_{xzx} - T_{zxx} - T_{xxz} - T_{yzy} - T_{zyy} - T_{yyz} + 2T_{zzz}\right)$$

$$\{3\}\mathrm{P}(0) = \left(-1/\sqrt{15}\right)\left(T_{xzx} + T_{zxx} + T_{xxz} + T_{xyzy} + T_{zyy} + T_{yyz} + 3T_{zzz}\right)$$

$$\{21\}'\mathrm{P}(0) = \left(-1/2\sqrt{3}\right)\left(T_{zxx} - 2T_{xxz} + T_{xzx} + T_{zyy} - 2T_{yyz} + T_{yzy}\right)$$

$$\{21\}''\mathrm{P}(0) = \left(-1/2\sqrt{3}\right)\left(T_{zxx} - 2T_{xzx} + T_{xxz} + T_{zyy} - 2T_{yzy} + T_{yyz}\right)$$

$$\{21\}'\overline{\mathrm{D}}(0) = (i/2)\left(T_{zxy} - T_{zyx} + T_{xzy} - T_{yzx}\right)$$

$$\{21\}''\overline{\mathrm{D}}(0) = (i/2)\left(T_{zxy} - T_{zyx} + T_{yxz} - T_{xyz}\right)$$

$$\{1^3\}\overline{\mathrm{S}} = \left(-i/\sqrt{6}\right)\left(T_{xyz} - T_{xzy} + T_{yzx} - T_{yxz} + T_{zxy} - T_{zyx}\right)$$

To conclude this part, it must be recalled that the norm $\left\|t^{\tau J}\right\|$ of an irreducible tensor $t^{\tau J}$ is an important notion since it allows one to compare, for some property, the importance of a term of given symmetry in two different compounds or to compare the relative importance of the terms of various symmetries in the same compound (examples in the field of nonlinear optics are given in References 7 and 8). In spherical coordinates $\left\|t^{\tau J}\right\|^2 = \sum_M t_M^{\tau J} t_M^{\tau J} *$; the norm of the original tensor $T$ is given by $\left\|T\right\|^2 = \sum_{ijk\ldots} T_{ijk\ldots} T_{ijk\ldots} *$ and, of course, one has $\left\|T\right\|^2 = \sum_{\tau J} \left\|t^{\tau J}\right\|^2$. The molecular selection rules in spherical coordinates for the most frequently encountered point groups are given for example in Reference 7.

## EXAMPLES

We will now briefly examine four well-known optical effects, namely linear susceptibility, optical activity, magneto-optical effects, and nonlinear susceptibilities; we will assume that a dielectric medium undergoes an optical field described by the vector potential

$$\mathbf{A} = \sum_{\mathbf{K}} \mathbf{e} a exp\ i(\mathbf{K}.\mathbf{r} - \omega t) + \mathbf{e}^* a^+ exp - i\ (\mathbf{K}.\mathbf{r} - \omega t)$$

($\omega$: frequency, $\mathbf{K}$ *:* wave-vector, $\mathbf{e}$ *:* polarization).

### Linear Susceptibility

In the electric dipole approximation, the linear polarizability tensor $\alpha(-\omega,\ \omega)$ and the linear susceptibility tensor $\chi(-\omega,\ \omega)$ are rank-2 tensors symmetric in the permutation of their indices from a general property of the kinetic coefficients; the light-matter scattering amplitude U is proportional to $\left(\mathbf{e}_s.\alpha.\mathbf{e}_i\right)$ where $\mathbf{e}_s$ and $\mathbf{e}_i$ are the polarization vectors of the incident and scattered fields respectively. Reversing the direction of time exchanges the roles of $\mathbf{e}_s$ and $\mathbf{e}_i$ but does not change U*,* hence $\alpha_{ij} = \alpha_{ji}$ *;* as a consequence there is no vector-type contribution to the linear susceptibility. Since $\alpha_{ij}$ may be complex, $\alpha_{ij} = \alpha_{ji}$ means that $\alpha$ is not hermitic; however the fields are real, which implies that $\alpha_{ij}(-\omega,\ \omega) = \alpha_{ij}{}^*\ (\omega, -\omega)$. The real and imaginary parts of $\alpha$*,* which account for dispersion and absorption in the medium respectively, are connected by the well-known Kramers—Kronig relations[9]. If the losses in the medium are negligible, $\alpha$ is invariant in the simultaneous permutation of the indices and of the frequencies, i.e. $\alpha_{ij}(-\omega,\ \omega) = \alpha_{ji}(\omega, -\omega)$; combining the latter equality with the former one, one finds $\alpha_{ij} = \alpha_{ji}{}^*$ that is $\alpha$ is hermitic. The losses are thus taken into account by the antihermitic part of $\alpha$.

A quantum-mechanical calculation provides the explicit expression of $\alpha$:

$$\alpha_{ij} = \sum_n \frac{\langle 0|d_i|n\rangle\langle n|d_j|0\rangle}{D_R} + \frac{\langle 0|d_j|n\rangle\langle n|d_i|0\rangle}{D_{NR}}$$

($\mathbf{d}$: electric dipole) with obvious notations and

$$D_R\left(resonant\ term\right) = E_n - E_0 - \delta_n - i\gamma_n/2 - \hbar\omega$$

$$D_{NR}\left(antiresonant\ term\right) = E_n - E_0 - \delta_n + i\gamma_n/2 + \hbar\omega$$

In Table 2 the irreducible spherical components of $\alpha$ are expressed as combinations of the original cartesian components.

**Optical Activity**

Optical activity arises from the dependence of the dielectric polarization $\mathbf{P}$ upon the spatial variation of the electromagnetic field

$$\mathbf{P} = 4\pi\chi\mathbf{E} + \eta : \nabla \otimes \mathbf{E} + ...$$

and is described by a rank-3 cartesian tensor $\eta_{ijk}$; the light-matter scattering amplitude U is then proportional to $\left(\mathbf{e}_s.\eta : i\mathbf{K} \otimes \mathbf{e}_i\right)$; reversing the direction of time changes that expression into $\left(-i\mathbf{K} \otimes \mathbf{e}_i{}^*.\eta : \mathbf{e}_s\right)$ without changing U, hence $\eta$ is anti-symmetric in the exchange of its extreme indices *i* and *k*, i.e. $\eta_{ijk} = -\eta_{kji}$.

A quantum mechanical calculation of $\eta$ based on a multipolar expansion of the field-matter interaction hamiltonian in the radiation gauge, i.e.

$$H = -\mathbf{d}.\mathbf{E} - \mathbf{M}.\mathbf{B} - \mathbf{Q} : \nabla \otimes \mathbf{E} + ...$$

($\mathbf{d}$: electric dipole, $\mathbf{M}$: magnetic dipole, $\mathbf{Q}$: electric quadrupole, $\mathbf{E}$: electric field, $\mathbf{B}$: magnetic field) shows both a magnetic dipole contribution $\eta^{(M)}$ and an electric quadrupole contribution $\eta^{(Q)}$ to the effect. The resulting expression for $\eta$ is:

$$\eta_{ijk} = \sum_n \frac{\langle 0|d_j|n\rangle\langle n|\tilde{Q}_{ik}|0\rangle + \langle 0|\tilde{Q}_{jk}|n\rangle\langle n|d_i|0\rangle}{D_{NR}} + \frac{\langle 0|d_i|n\rangle\langle n|\tilde{Q}_{jk}|0\rangle + \langle 0|\tilde{Q}_{ik}|n\rangle\langle n|d_j|0\rangle}{D_R}$$

where $\tilde{\mathbf{Q}}$ is the hermitic tensor operator defined by

$$\tilde{Q}_{ij} = Q_{ij} - \frac{i}{\omega}\sum_k \varepsilon_{ijk} M_k$$

The magnetic dipole contribution is a true rank-3 tensor anti-symmetric in the exchange of its extreme indices; since $\mathbf{M}$ is a pseudo-vector ($\overline{\mathrm{P}}$ type), $\eta^{(M)}$ has three irreducible parts, a pseudo-quadrupole $\{21\}\overline{\mathrm{D}}$, a true vector $\{21\}\mathrm{P}$, and a pseudo-scalar $\{1^3\}\overline{\mathrm{S}}$.

The electric quadrupole contribution is also a true rank-3 cartesian tensor anti-symmetric in the exchange of its extreme indices; since $\mathbf{Q}$ is a symmetric and traceless rank-2 tensor ($\mathrm{D}$ type), $\eta^{(Q)}$ has only two irreducible parts, a pseudo-quadrupole $\{21\}\overline{\mathrm{D}}$ and a true vector $\{21\}\mathrm{P}$. When comparing the reduction spectra of $\eta^{(M)}$ and $\eta^{(Q)}$, one sees that $\eta$ can be written as $\overline{\mathrm{D}} + \mathrm{P} + \overline{\mathrm{S}}$; that is the reduction spectrum of a cartesian rank-2 pseudo-tensor; optical activity can thus be described by a pseudo-tensor $\overline{g}_{jl} = \sum_{pq} \varepsilon_{pjq}\eta_{plq}$ , and the dielectric polarization can be written as

$$\mathbf{P} = 4\pi\chi.\mathbf{E} + (\overline{g}.\nabla)\,.\mathbf{E} + \ldots$$

In an isotropic or cubic medium where the pseudo quadrupole and the vector vanish, optical activity only arises from the pseudo-scalar term, which is entirely due to the magnetic dipole contribution; to our knowledge, the vector-type term, which is presumably very weak since it is the antisymmetric part of $\overline{g}$, has not been observed yet.

In Table 3 the irreducible spherical components of $\eta$ are expressed as combinations of the original cartesian components; the symmetry condition $\eta_{ijk} = -\eta_{kji}$ involves the irreducible parts $\{21\}\mathrm{P}''$ and $\{21\}\overline{\mathrm{D}}''$ vanishing.

**Magneto-Optical Effects**

In the presence of a strong static magnetic field, the dielectric polarization $\mathbf{P}$ at the frequency w can be expanded as a series of the applied magnetic field $\mathbf{B}$

$$\mathbf{P} = 4\pi\chi.\mathbf{E} + \frac{1}{2!}\tilde{\beta}:\mathbf{BE} - \frac{1}{3!}\ \tilde{\gamma}::\mathbf{BBE} + ...$$

In that expansion, $\tilde{\beta}$ is a pseudo rank-3 cartesian tensor; the light–matter scattering amplitude is proportional to $\left(\mathbf{e}_s *.\ \tilde{\beta}:\mathbf{B}\ \mathbf{e}_i\right)$; reversing the direction of time changes that expression into $\left(\mathbf{e}_i *.\ -\tilde{\beta}:\mathbf{B}\ \mathbf{e}_s\right)$ without changing U, hence $\tilde{\beta}$ is anti-symmetric in the exchange of its extreme indices; that is a general property of the kinetic coefficient in the presence of a magnetic field. $\tilde{\beta}$ accounts for circular birefringence (Faraday effect); its irreducible parts are $\{21\}\mathrm{D}, \{21\}\overline{\mathrm{P}}, \{1^3\}\mathrm{S}$ so that it transforms like a rank-2 true tensor $g$ so that the dielectric polarization is

$$\mathbf{P} = \left(4\pi\chi - \frac{1}{3!}\ \tilde{\gamma}::\mathbf{BB}\right).\mathbf{E} + \frac{1}{2!}\left(g.\mathbf{B}\right).\mathbf{E} + ...$$

$\tilde{\gamma}$ is a rank-4 true tensor, symmetric in the permutation of its two extreme coefficients (a property of the kinetic coefficients) or its two middle indices (due to the indiscernability of the two magnetic fields), and accounts for quadratic magnetic birefringence (Cotton–Mouton effect); its reduction spectrum is $\{4\}(\mathrm{S} + \mathrm{D} + \mathrm{G}), \{2^2\}(\mathrm{S} + \mathrm{D})$.

## NONLINEAR OPTICS

The (r-1)$^{th}$ order dielectric polarization at the frequency $\omega_1$ can be expanded as a series of the (r-1) incident electric fields at the respective frequencies $\omega_2, \omega_3, ..., \omega_r$

$$\mathbf{P}^{(r-1)}(\omega_1) = \chi^{(r)}(-\omega_1, \omega_2, \omega_3, ..., \omega_r)\mathbf{E}(\omega_2)\mathbf{E}(\omega_3)...\mathbf{E}(\omega_r)$$

The susceptibility $\chi^{(r)}(-\omega_1, \omega_2, \omega_3, ..., \omega_r)$ which describes the mixing of *r* frequencies $\omega_1, \omega_2, \omega_3, ..., \omega_r$ such that $\omega_1 = \omega_2 + \omega_3 + ... + \omega_r$ is a rank-r cartesian true tensor with the following properties[10:]

(i) the reality of the fields implies that $\chi^{(r)}$ is changed into $\chi^{(r)*}$` if one changes the sign of all the frequencies;

(ii) if there are no losses in the medium, $\chi^{(r)}$ is invariant in any simultaneous permutation of the indices and also of the frequencies.

Moreover, if the dispersion is weak the susceptibility tensor must approximately be symmetric in any index permutation. Thus, in the transparency range of the medium, the dominant parts of the reduction spectrum are those which appear in the spectrum of a fully symmetric tensor of the same rank while the other ones are, presumably, small. As a consequence, in the general case, the fully symmetric parts include resonant and non-resonant terms whereas the other ones only contain resonant terms.

### Three-Wave Mixing

The first hyperpolarizability tensor $\beta(-\omega_1, \omega_2, \omega_3)$ and the first nonlinear susceptibility tensor $\chi^{(2)}(-\omega_1, \omega_2, \omega_3)$ describe the mixing of three frequencies. For instance, the fully antisymmetric pseudoscalar

$$\{1^3\}\bar{\mathrm{S}} = \left(\frac{-i}{\sqrt{6}}\right)\left(\chi^{(2)}_{xyz} - \chi^{(2)}_{xzy} + \chi^{(2)}_{yzx} - \chi^{(2)}_{yxz} + \chi^{(2)}_{zxy} - \chi^{(2)}_{zyx}\right)$$

accounts for the mixing of three different frequencies in an optically active isotropic medium such as a solution of chiral molecules.

The electro-optical effect is described by the tensor $\chi^{(2)}(-\omega,0,\omega)$, which is symmetrical in the permutation of its extreme indices from a time-reversal argument (unlike the magnetic field in the magneto-optic effect, the electric field is unchanged); the reduction spectrum of $\chi^{(2)}(-\omega,0,\omega)$ thus reduces to $\{3\}(\mathrm{P}+\mathrm{F}),\{21\}(\mathrm{P}+\overline{\mathrm{D}})$.

The optical rectification effect is described by the tensor $\chi^{(2)}(0,-\omega,\omega)$ which is invariant in the permutation of its two last indices from a time-reversal argument and shows a similar reduction spectrum. The second-harmonic generation effect is described by the tensor $\chi^{(2)}(-2\omega,\omega,\omega)$ which is symmetric in the permutation of its two last indices because of the indiscernability of the two fundamental fields and again shows the same reduction spectrum, i.e. $\{3\}(\mathrm{P}+\mathrm{F})$, $\{21\}(\mathrm{P}+\overline{\mathrm{D}})$.

In Table 4 the irreducible spherical components of $\beta(-2\omega,\omega,\omega)$ are expressed as combinations of the original cartesian components; the symmetry condition $\beta_{ijk}=\beta_{ikj}$ implies that the irreducible parts $\{21\}\,\mathrm{P}'$ and $\{21\}\,\mathrm{P}''$ are identical along with $\{21\}\overline{\mathrm{D}}'$ and $\{21\}\overline{\mathrm{D}}''$.

## Four-Wave Mixing

The second hyperpolarizability tensor $\gamma(-\omega_1,\omega_2,\omega_3,\omega_4)$ and the second nonlinear susceptibility tensor $\chi^{(3)}(-\omega_1,\omega_2,\omega_3,\omega_4)$ describe the mixing of four frequencies. For instance the coherent Raman effect tensor $\chi^{(3)}(-2\omega_1\pm\omega_2,\mp\omega_2,\omega_1,\omega_1)$ is symmetric in the permutation of its two last indices, so that its reduction spectrum is $\{4\}(\mathrm{S}+\mathrm{D}+\mathrm{G})$, $\{31\}2(\overline{\mathrm{P}}+\mathrm{D}+\overline{\mathrm{F}})$, $\{2^2\}(\mathrm{S}+\mathrm{D})$, $\{21^2)\overline{\mathrm{P}}$; the third-harmonic generation tensor $\chi^{(3)}(-3\omega,\omega,\omega,\omega)$ is fully symmetric in the permutation of its three last indices so that its reduction spectrum reduces to $\{4\}(\mathrm{S}+\mathrm{D}+\mathrm{G})$, $\{31\}(\overline{\mathrm{P}}+\mathrm{D}+\overline{\mathrm{F}})$. We will ultimately consider the optical Kerr effect, which is described by the tensor $\chi^{(3)}(-\omega,\omega_p,-\omega_p,\omega)$; $\omega_p$ is the frequency of the pump beam, $\omega$ the frequency of the signal beam; in the static Kerr effect, $\omega_p=0$. $\chi^{(3)}(-\omega,\omega_p,-\omega_p,\omega)$ is symmetric in the permutation of its two extrema and of its two middle indices; its reduction spectrum thus reduces to $\{4\}(\mathrm{S+D+G})$, $\{2^2\}(\mathrm{S+D})$, and ultimately, in the degenerate configuration $\omega=\omega_p$, to the fully symmetric term $\{4\}(\mathrm{S}+\mathrm{D}+\mathrm{G})$. In the non-degenerate configuration the case of an isotropic medium will be considered; the reduction spectrum then reduces to two scalars, one totally symmetric $\{4\}\mathrm{S}$ and one non-totally-symmetric $\{2^2\}\mathrm{S}$; with Young's methods, one can find the normalized expressions of those two scalars:

$$\{4\}\mathrm{S}=\left(1/\sqrt{5}\right)\left(3\chi^{(3)}_{xxxx}+2\chi^{(3)}_{xxyy}+2\chi^{(3)}_{xyxy}+2\chi^{(3)}_{xyyx}\right)$$

$$\{2^2\}\mathrm{S}=\left(2\chi^{(3)}_{xxyy}-\chi^{(3)}_{xyyx}-\chi^{(3)}_{xyxy}\right)$$

Let $\mathbf{e}_p$, $\mathbf{e}_i$, $\mathbf{e}_s$ be the polarization vectors of the pump beam, of the incident signal beam, and of the analysed signal beam respectively. The scattered amplitude is proportional to

$$\mathrm{U}=\chi^{(3)}(-\omega,\omega_p,-\omega_p,\omega)\ ::\ \mathbf{e}_s\mathbf{e}_p\mathbf{e}_p{}^*\mathbf{e}_i$$

In terms of irreducible tensors, this expression is changed into

$$\mathrm{U}=\sum_{\tau J}\chi^{\tau J}\cdot\xi^{\tau J}\quad\text{with } \boldsymbol{\xi}=\mathbf{e}_s\otimes\mathbf{e}_p\otimes\mathbf{e}_p{}^*\otimes\mathbf{e}_i$$

Since only the isotropic parts (J = 0) of $\chi^{(3)}$ do not vanish, one deduces the following selection rules: if all the $\mathbf{e}$ are parallel to each other, one can only observe the fully symmetric part of $\chi^{(3)}$; conversely if $\mathbf{e}_p$ is circularly polarized and if $\mathbf{e}_i$ and $\mathbf{e}_s$ are linearly cross-polarized, one will only detect the non-fully-symmetric term.

## CONCLUSION

We have emphasized the importance of index permutations in the reduction of cartesian tensors with respect to the orthogonal group and its subgroups. First the intrinsic symmetries of the physical effect under investigation must be considered; these symmetries are associated in the representative tensors with some index permutation symmetries which cause some of the tensor parts, those that are irreducible under the permutation group, to vanish. The remaining parts can be reduced under the orthogonal group. Taking into account the effects of point-group invariances upon tensors constitutes the ultimate step of the process.

## ACKNOWLEDGEMENT

The author feels indebted to J. Zyss (CNET) for fruitful discussions on the manuscript.

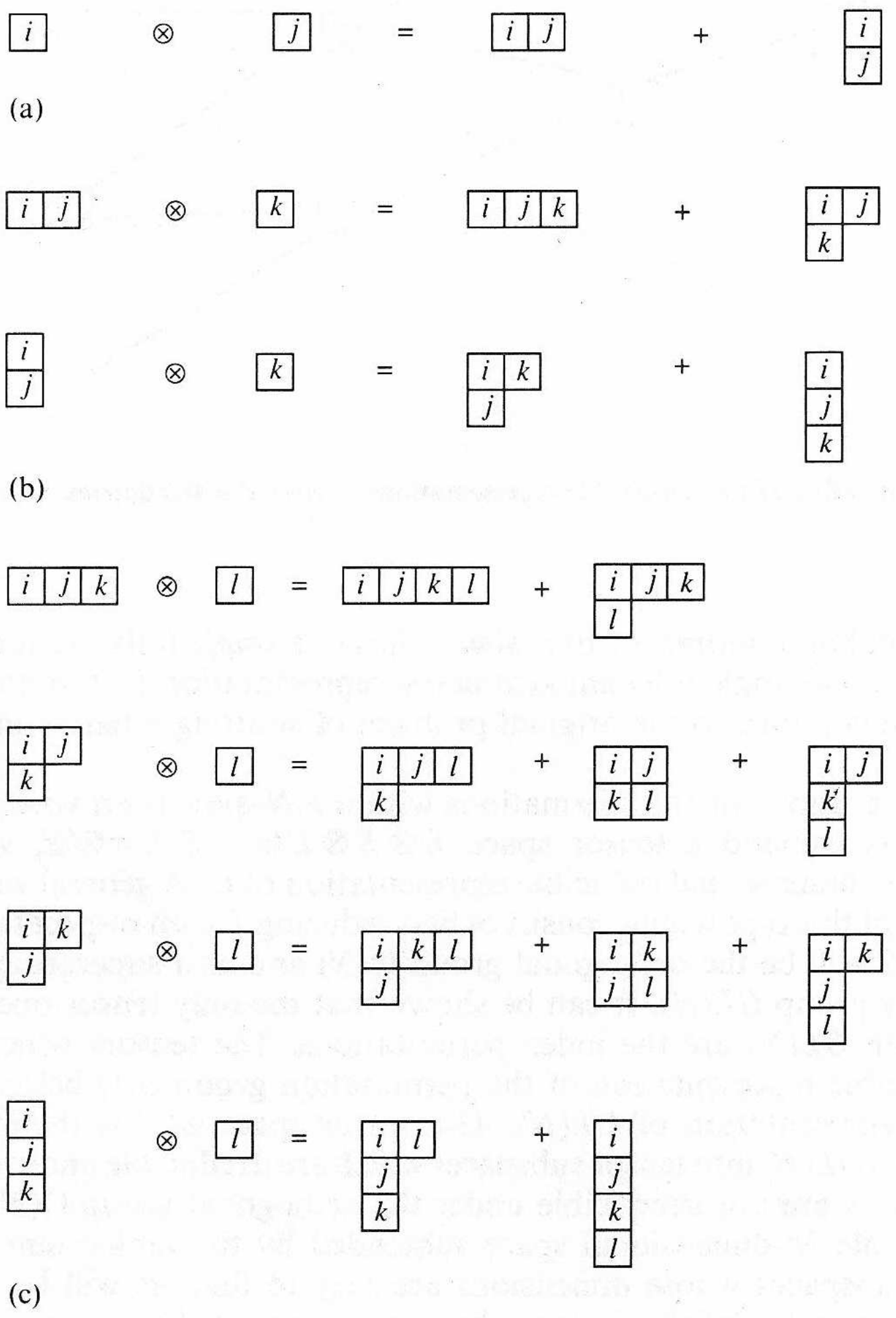

FIGURE 1 Construction of the basis vectors of the irreducible representations of $\sigma(r)$ from those of $\sigma(r-1)$: a) $r = 2$ (2); b) $r = 3$; c) $r = 4$.

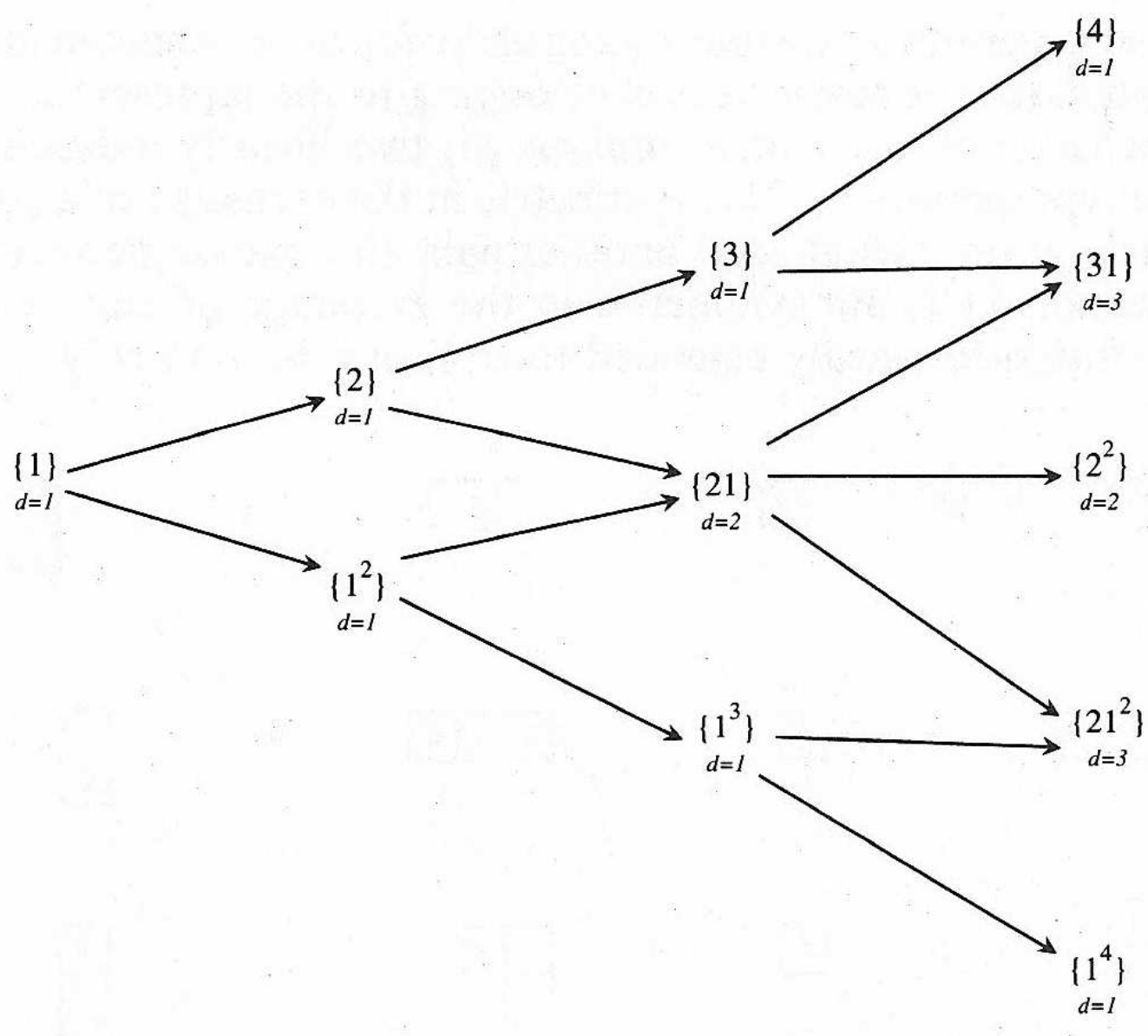

FIGURE 2 Generation of the irreducible representations of $\sigma(r)$; $d$ is the dimension of the representation.

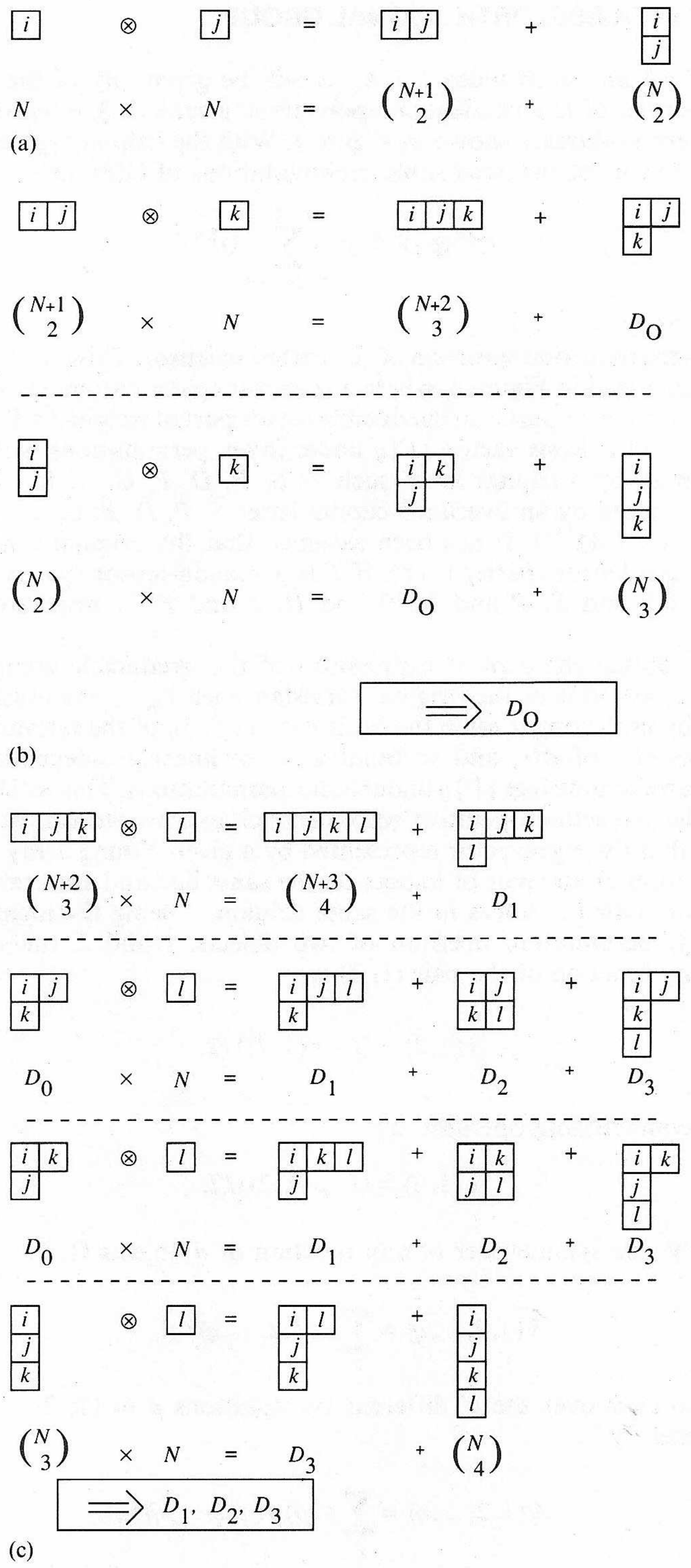

FIGURE 3 Determination of the number of independent components of a rank-$r$ tensor irreducible under the index permutation group $(r)$: a) $r = 2$; b) $r = 3$; c) $r = 4$.

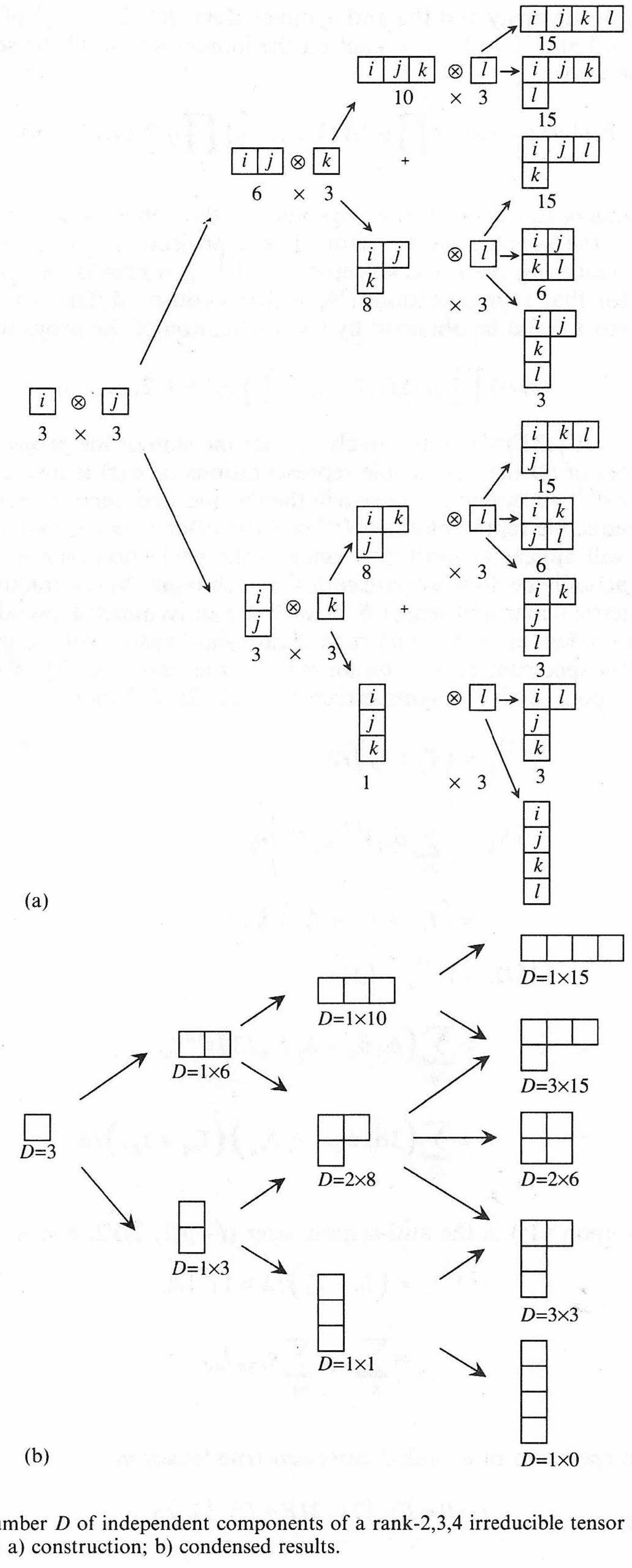

FIGURE 4 Number $D$ of independent components of a rank-2,3,4 irreducible tensor in the three-dimension case: a) construction; b) condensed results.

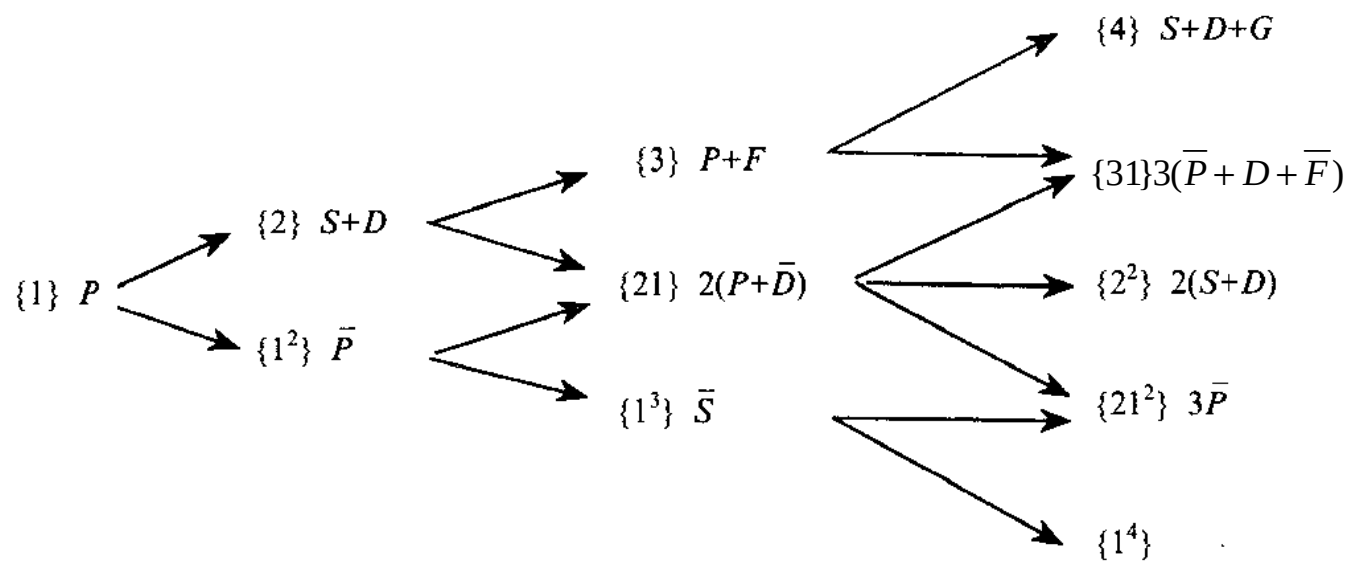

FIGURE 5 Reduction spectrum of a true cartesian tensor under index permutations and orthogonal transformations.

TABLE 1

Explicit expression of the reduced spherical harmonics $C_M^J$ in terms of the components $u_x$, $u_y$, $u_z$ of a unit vector in the direction of polar angles $\theta,\varphi$

$$C^0 = 1$$
$$C_0^1 = u_z$$
$$C_{\pm 1}^1 = \mp (1/2)^{1/2}(u_x \pm i u_y)$$
$$C_0^2 = (1/2)\left(2u_z^2 - u_x^2 - u_y^2\right)$$
$$C_{\pm 1}^2 = \mp (3/2)^{1/2}(u_z u_x \pm i u_z u_y)$$
$$C_{\pm 2}^2 = (3/8)^{1/2}\left(u_x^2 - u_y^2 \pm 2i u_x u_y\right)$$
$$C_0^3 = (1/2)(2u_y^3 - 3u_z u_x^3 - 3u_z u_y^2)$$
$$C_{\pm 1}^3 = \mp (3/16)^{1/2}(4u_z^2 u_x - u_x^3 - u_x u_y^2 \pm 4 i u_z^2 u_y \mp i u_x u_y \mp i u_y^3)$$
$$C_{\pm 2}^3 = (15/8)^{1/2}(u_z u_x^2 - u_z u_y^2 \pm 2i u_z u_x u_y)$$
$$C_{\pm 3}^3 = \mp (5/16)^{1/2}(u_x^3 \pm 3i u_x^2 u_y - 3 u_x u_y^2 \mp i u_y^3)$$

TABLE 2

Irreducible spherical components of the linear polarizability tensor $\alpha(-\omega, \omega)$:

$$\alpha_0^{(0)} = -\frac{1}{\sqrt{3}}(\alpha_{xx} + \alpha_{yy} + \alpha_{zz})$$

$$\alpha_0^{(1)} = \frac{i}{\sqrt{2}}(\alpha_{xy} - \alpha_{yx})$$

$$\alpha_{\pm1}^{(1)} = \mp\frac{i}{\sqrt{2}}(\alpha_{yz} - \alpha_{zy} \pm i\,\alpha_{zx} \mp i\,\alpha_{xz})$$

$$\alpha_0^{(2)} = \frac{1}{\sqrt{6}}(2\alpha_{zz} - \alpha_{xx} - \alpha_{yy})$$

$$\alpha_{\pm1}^{(2)} = \mp\frac{1}{2}(\alpha_{zx} + \alpha_{xz} \pm i\,\alpha_{zy} \pm i\,\alpha_{yz})$$

$$\alpha_{\pm2}^{(2)} = \frac{1}{2}(\alpha_{xx} - \alpha_{yy} \pm i\,\alpha_{xy} \pm i\,\alpha_{yx})$$

TABLE 3

Irreducible spherical components of the optical activity tensor $\eta$:

$$\eta_0^{(0)} = \frac{-2i}{\sqrt{6}}(\eta_{xyz} + \eta_{yzx} + \eta_{zxy})$$

$$\eta_0^{(1)} = -\frac{\sqrt{3}}{2}(\eta_{zxx} + \eta_{zyy})$$

$$\eta_{\pm1}^{(1)} = \pm\frac{\sqrt{6}}{4}(\eta_{xyy} + \eta_{xzz}) + \frac{i\sqrt{6}}{4}(\eta_{yxx} + \eta_{yzz})$$

$$\eta_0^{(2)} = \frac{i}{2}(\eta_{xyz} + \eta_{zxy} - 2\eta_{yzx})$$

$$\eta_{\pm1}^{(2)} = \frac{i\sqrt{3}}{2}(\eta_{yxx} + \eta_{zzy}) - \frac{\sqrt{3}}{2}(\eta_{yyx} + \eta_{xxz})$$

$$\eta_{\pm2}^{(2)} = \frac{i\sqrt{3}}{2}(\eta_{xyz} - \eta_{zxy}) \mp \frac{\sqrt{3}}{2}(\eta_{zxx} - \eta_{zyy})$$

TABLE 4

Irreducible spherical components of the nonlinear polarizability tensor $\beta\,(-2\omega,\omega,\omega)$ accounting for second-harmonic generation

$$\{3\}\,\beta^{(3)}_{0} = \frac{1}{\sqrt{10}}\,(-\beta_{zxx} - \beta_{xzx} - \beta_{xxz} - \beta_{zyy} - \beta_{yzy} - \beta_{yyz} + 2\beta_{zzz})$$

$$\{3\}\,\beta^{(3)}_{\pm 1} = \pm\frac{1}{2\sqrt{30}}\,(3\beta_{xxx} + \beta_{xyy} + \beta_{yxy} + \beta_{yyx} - 4\beta_{xzz} - 4\beta_{zxz} - 4\beta_{zzx}) + \frac{i}{2\sqrt{30}}\,(3\beta_{yyy} + \beta_{yxx} + \beta_{xyx} + \beta_{xxy} - 4\beta_{yzz} - 4\beta_{zyz} - 4\beta_{zzy})$$

$$\{3\}\,\beta^{(3)}_{\pm 2} = \frac{1}{2\sqrt{3}}\,(\beta_{zxx} + \beta_{xzx} + \beta_{xxz} - \beta_{zyy} - \beta_{yzy} - \beta_{yyz}) \pm \frac{i}{2\sqrt{3}}\,(\beta_{xyz} + \beta_{xzy} + \beta_{yzx} + \beta_{yxz} + \beta_{zxy} + \beta_{zyx})$$

$$\{3\}\,\beta^{(3)}_{\pm 3} = \pm\frac{1}{2\sqrt{2}}\,(-\beta_{xxx} + \beta_{xyy} + \beta_{yxy} + \beta_{yyx}) - \frac{i}{2\sqrt{2}}\,(-\beta_{yyy} + \beta_{yxx} + \beta_{xyx} + \beta_{xxy})$$

$$\{3\}\,\beta^{(1)}_{0} = \frac{1}{\sqrt{15}}\,(-\beta_{zxx} - \beta_{xzx} - \beta_{xxz} - \beta_{zyy} - \beta_{yzy} - \beta_{yyz} - 3\beta_{zzz})$$

$$\{3\}\,\beta^{(1)}_{\pm 1} = \pm\frac{1}{\sqrt{30}}\,(3\beta_{xxx} + \beta_{xyy} + \beta_{yxy} + \beta_{yyx} + \beta_{xzz} + \beta_{xzx} + \beta_{zzx}) + \frac{i}{\sqrt{30}}\,(\beta_{yyy} + \beta_{yxx} + \beta_{xyx} + \beta_{xxy} + \beta_{yzz} + \beta_{zyz} + \beta_{zzy})$$

$$\{21\}\,\beta^{(1)}_{0} = -\frac{1}{\sqrt{3}}\,(\beta_{zxx} - \beta_{xxz} + \beta_{zyy} - \beta_{yyz})$$

$$\{21\}\,\beta^{(1)}_{\pm 1} = \pm\frac{1}{\sqrt{6}}\,(\beta_{xyy} - \beta_{yyx} + \beta_{xzz} - \beta_{zzx}) + \frac{i}{\sqrt{6}}\,(\beta_{yxx} - \beta_{xxy} + \beta_{yzz} - \beta_{zzy})$$

$$\{21\}\,\beta^{(2)}_{0} = i(\beta_{xyz} - \beta_{yzx})$$

$$\{21\}\,\beta^{(2)}_{\pm 1} = \mp\frac{i}{\sqrt{2}}\,(\beta_{xxy} - \beta_{yxx}) + \frac{1}{\sqrt{2}}\,(\beta_{xyy} - \beta_{yyx})$$

$$\{21\}\,\beta^{(2)}_{\pm 2} = \frac{i}{2}\,(\beta_{yzx} - \beta_{zxy} + \beta_{xyz} - \beta_{zyx}) \mp \frac{1}{2}\,(\beta_{yyz} - \beta_{zyy} + \beta_{zxx} - \beta_{xxz})$$